\documentstyle[12pt,epsfig,graphics,cite,amssymb,amsmath,bm]{article}
\begin{document}\bibliographystyle{plain}\begin{titlepage}
\renewcommand{\thefootnote}{\fnsymbol{footnote}}\hfill
\begin{tabular}{l}HEPHY-PUB 952/15\\UWThPh-2015-20\\
September 2015\end{tabular}\\[2cm]\Large\begin{center}{\bf LIGHT
PSEUDOSCALAR MESONS IN BETHE--SALPETER EQUATION WITH INSTANTANEOUS
INTERACTION}\\[1cm]\large{\bf Wolfgang
LUCHA\footnote[1]{\normalsize\ {\em E-mail address\/}:
wolfgang.lucha@oeaw.ac.at}}\\[.3cm]\normalsize Institute for High
Energy Physics,\\Austrian Academy of Sciences,\\Nikolsdorfergasse
18, A-1050 Vienna, Austria\\[1cm]\large{\bf Franz
F.~SCH\"OBERL\footnote[2]{\normalsize\ {\em E-mail address\/}:
franz.schoeberl@univie.ac.at}}\\[.3cm]\normalsize Faculty of
Physics, University of Vienna,\\Boltzmanngasse 5, A-1090 Vienna,
Austria\\[2cm]{\normalsize\bf Abstract}\end{center}\normalsize

\noindent The light pseudoscalar mesons play a twofold r\^ole:
they may or have to be regarded both as low-lying bound states of
the fundamental degrees of freedom of quantum chromodynamics as
well as the (pseudo-) Goldstone bosons of the spontaneously broken
chiral symmetries of quantum chromodynamics. We interrelate these
aspects in a single quantum-field-theoretic approach relying on
the Bethe--Salpeter formalism in instantaneous approximation by
very simple means: the shape of the pseudoscalar-meson
Bethe--Salpeter wave function dictated by chiral symmetry is used
in Bethe--Salpeter equations for bound states of vanishing mass,
in order to deduce analytically the interactions which govern the
bound states under study. In this way, we obtain exact
Bethe--Salpeter solutions for pseudoscalar mesons, in the sense of
establishing the rigorous relationship between, on the one hand,
the relevant~interactions and, on the other hand, the
Bethe--Salpeter amplitudes that~characterize the bound states.
\vspace{3ex}

\renewcommand{\thefootnote}{\arabic{footnote}}\end{titlepage}

\section{Introduction}Light pseudoscalar mesons may be understood
as bound states of a quark and an antiquark but they have to be
interpreted also as (almost) massless (pseudo-) Goldstone bosons
of the spontaneously broken chiral symmetries of quantum
chromodynamics, the theory of strong interactions, whence their
description still poses a challenge to theoretical particle
physics. Within quantum field theory, the Bethe--Salpeter
framework provides a Poincar\'e-covariant approach to bound states
\cite{BSE}. Some of its inherent obstacles can be avoided by
restriction to three-dimensional, e.g.\ instantaneous, reductions,
such as Salpeter's equation \cite{SE}. Recently, by inversion of
this bound-state problem, we showed that, under favourable
circumstances, the underlying interaction potential can be
retrieved from the Salpeter solutions \cite{WL13}. In~this paper,
we apply the inversion technique of Ref.~\cite{WL13}\footnote{In
Ref.~\cite{WL13}, we focused (mainly) to the construction of exact
solutions to the {\em reduced\/} Salpeter equation. The reduced
Salpeter equation emerges from the (full) Salpeter equation when
assuming the systems~under consideration to be composed of weakly
bound semirelativistic heavy constituents. In contrast, the
present analysis deals with the {\em full\/} Salpeter equation,
which does not need any such requirement. Hence, the rather severe
constraints on the nature of the bound states do not apply in the
case of the (full) Salpeter~equation.} to pseudoscalar
Goldstone-type mesons (like the pion), to see how the strong
interactions enter such kind of bound-state equation.\footnote{We
would like to convey in this way our gratitude to the referee of
Ref.~\cite{WL13} for instigating these~analyses by expressing
great interest in the application of the formalism elaborated in
Ref.~\cite{WL13} to the case of the~pion.}

The outline of this paper is as follows. In Sec.~\ref{Sec:FIBSE},
we recall briefly and thus in a~somewhat symbolic notation only
those aspects of the Bethe--Salpeter formalism that are of
relevance for the subsequent discussion. In Sec.~\ref{Sec:IP}, we
sketch the, in fact, not particularly complicated concepts behind
our inversion procedures. In Sec.~\ref{Sec:PSA}, in an attempt to
get the most out~of~it, we squeeze dry the relevant results
emerging from Euclidean-space-based analyses utilizing the
Dyson--Schwinger framework, in order to extract from these
well-motivated conjectures how the bound-state amplitudes forming
the starting point of the inversion might look like. Armed with
these insights, it is just one small step for (a) man to recover,
in Sec.~\ref{Sec:CSP}, for both zero (Subsec.~\ref{sec:m=0}) and
non-zero (Subsec.~\ref{sec:m>0}) equal mass of the two
(anti-)~quarks bound to Goldstone bosons the basic interactions.
Finally, Sec.~\ref{Sec:SCO} is devoted to summarizing~remarks.

\section{Bethe--Salpeter Formalism in Instantaneous Limit}
\label{Sec:FIBSE}An inevitable prerequisite of the application of
our inversion approach developed in Ref.~\cite{WL13} to
quark--antiquark bound states of Goldstone type is clearly the
sufficient simplification of any quantum-field-theoretic
description of bound states. Let us thus recall its crucial~steps.

The Bethe--Salpeter formalism \cite{BSE} describes the features of
bound states of fundamental degrees of freedom of one's quantum
field theory in use by the solutions of the \emph{homogeneous
Bethe--Salpeter equation}, deduced in form of pairs ($M,$ $\Phi$)
of its (discrete) bound-state mass eigenvalues $M$ and associated
eigenstates, represented by Bethe--Salpeter amplitudes $\Phi.$ In
momentum-space representation, each such Bethe--Salpeter amplitude
$\Phi(p,P)$ encodes the distribution of the relative momenta $p$
of the two constituents of the respective bound~state of total
momentum $P.$ The Bethe--Salpeter equation relates this
Bethe--Salpeter amplitude $\Phi(p,P),$ for bound-state
constituents having momenta $p_i,$ $i=1,2,$ to their full
propagators $S_i(p_i)$ and an integral kernel $K(p,q,P)$ that
encompasses all interactions of these particles:
\begin{equation}\Phi(p,P)=\frac{\rm i}{(2\pi)^4}\,S_1(p_1)\int{\rm
d}^4q\,K(p,q,P)\,\Phi(q,P)\,S_2(-p_2)\ .\label{Eq:BSE}
\end{equation}

In order to render this Bethe--Salpeter formalism accessible to
our inversion techniques, we consider some so-called
three-dimensional reduction of this framework. The by far most
popular among such class of approximations is the static limit,
encountered if assuming the bound-state constituents, in their
center-of-momentum frame defined by $\bm{p}=\bm{p}_1=-\bm{p}_2,$
to interact instantaneously. The kernel then depends only on the
relative three-momenta~$\bm{p},\bm{q}$:
$$K(p,q,P)=K(\bm{p},\bm{q})\ .$$If the propagators $S_i(p_i)$ do
not involve any \emph{non-trivial\/} dependence on the time
component $p_0$ of the relative momentum $p,$ the Bethe--Salpeter
equation (\ref{Eq:BSE}) becomes, upon integration over $p_0,$ a
generalized \cite{WL05:IBSEWEP} instantaneous Bethe--Salpeter
equation for the Salpeter amplitude\begin{equation}
\phi(\mbox{\boldmath{$p$}})\equiv\frac{1}{2\pi}\int{\rm
d}p_0\,\Phi(p)\ .\label{Eq:SA}\end{equation}An example of such
kind of equation has been derived in
Refs.~\cite{WL05:IBSEWEP,WL06:GIBSEEQP-C7} and explored
in~Ref.~\cite{WL05:EQPIBSE}.

The easiest way to accomplish the desired triviality of the
$p_0$-dependence of any fermion propagator is to follow Salpeter
\cite{SE} by assuming each bound-state constituent to propagate
freely and, accordingly, approximating in Eq.~(\ref{Eq:BSE}) each
full propagator $S_i(p)$ by its free form $S_{i,0}(p,m_i),$ with,
however, an \emph{effective\/} mass $m_i$ subsuming all dynamical
self-energy~effects:$$S_{i,0}(p,m_i)=\frac{{\rm
i}}{\not\!p-m_i+{\rm i}\,\varepsilon}\equiv{\rm
i}\,\frac{\not\!p+m_i}{p^2-m_i^2+{\rm i}\,\varepsilon}\
,\qquad\not\!p\equiv p^\mu\,\gamma_\mu\
,\qquad\varepsilon\downarrow0\ ,\qquad i=1,2\ .$$Expressed in
terms of the one-particle energy $E_i(\bm{p}),$ one-particle Dirac
Hamiltonian $H_i(\bm{p}),$ and energy projection operators
$\Lambda_i^\pm(\bm{p})$ for positive and negative energy of
particle $i=1,2,$ $$E_i(\bm{p})\equiv\sqrt{\bm{p}^2+m_i^2}\
,\qquad H_i(\bm{p})\equiv\gamma_0\,(\bm{\gamma}\cdot\bm{p}+m_i)\
,\qquad\Lambda_i^\pm(\bm{p})\equiv\frac{E_i(\bm{p})\pm
H_i(\bm{p})}{2\,E_i(\bm{p})}\ ,$$the result of these
simplifications of the Bethe--Salpeter equation (\ref{Eq:BSE}) is
(upon application of contour integration in the complex-$p_0$
plane and residue theorem) the Salpeter equation \cite{SE}
\begin{align}\phi(\bm{p})&=\int\frac{{\rm d}^3q}{(2\pi)^3}
\left(\frac{\Lambda_1^+(\bm{p}_1)\,\gamma_0\,
[K(\bm{p},\bm{q})\,\phi(\bm{q})]\,\gamma_0\,\Lambda_2^-(\bm{p}_2)}
{P_0-E_1(\bm{p}_1)-E_2(\bm{p}_2)}\right.\nonumber\\[1ex]
&\hspace{11.11ex}\left.-\frac{\Lambda_1^-(\bm{p}_1)\,\gamma_0\,
[K(\bm{p},\bm{q})\,\phi(\bm{q})]\,\gamma_0\,\Lambda_2^+(\bm{p}_2)}
{P_0+E_1(\bm{p}_1)+E_2(\bm{p}_2)}\right).\label{Eq:SE}\end{align}
Its specific projector structure subjects the Salpeter amplitude
(\ref{Eq:SA}) to the crucial constraint
\begin{equation}\Lambda_1^+(\bm{p}_1)\,\phi(\bm{p})\,\Lambda_2^+(\bm{p}_2)=
\Lambda_1^-(\bm{p}_1)\,\phi(\bm{p})\,\Lambda_2^-(\bm{p}_2)=0\
.\label{Eq:SAC}\end{equation}

Counting the number of basis elements of the Dirac algebra, the
Salpeter amplitude can have, in principle, at most 16 independent
components. The constraints (\ref{Eq:SAC}), however, halve this
number: the most general Salpeter amplitude $\phi(\bm{p})$ has
eight independent components. The Salpeter amplitude of bound
states of a spin-$\frac{1}{2}$ fermion and a spin-$\frac{1}{2}$
antifermion~whose spin quantum numbers add up to zero has merely
two independent components, henceforth labelled
$\varphi_1(\bm{p})$ and $\varphi_2(\bm{p})$; pseudoscalar bound
states are just that special case of this where the relative
orbital angular momentum $\ell$ of these bound-state constituents
vanishes as well. The constraint (\ref{Eq:SAC}) enforces, as
general form of any such ($CP=-1$) Salpeter amplitude \cite{REE},
\begin{equation}\phi(\bm{p})
=\left[\varphi_1(\bm{p})\,\frac{H(\bm{p})}{E(\bm{p})}
+\varphi_2(\bm{p})\right]\gamma_5\ .\label{Eq:PSA}\end{equation}

The interaction kernel $K(\bm{p},\bm{q})$ in Salpeter's equation
can be written in form of a sum of products of tensor products
$\Gamma_1\otimes\Gamma_2$ of Dirac matrices $\Gamma_{1,2},$
defining the Lorentz structure of the \emph{effective\/} couplings
of the fermions, and associated Lorentz-scalar potentials
generically labelled $V(\bm{p},\bm{q}).$ Here, we find reasonable
to assume the equality $\Gamma_1=\Gamma_2=\Gamma$ of $\Gamma_1$
and~$\Gamma_2$:\begin{equation}[K(\bm{p},\bm{q})\,\phi(\bm{q})]=
\sum_\Gamma V_\Gamma(\bm{p},\bm{q})\,\Gamma\,\phi(\bm{q})\,\Gamma\
.\label{Eq:BSK}\end{equation}Convolution nature and spherical
symmetry of all functions
$V_\Gamma(\bm{p},\bm{q})=V_\Gamma((\bm{p}-\bm{q})^2)$ --- and
hence of the entire kernel $K(\bm{p},\bm{q})=K((\bm{p}-\bm{q})^2)$
--- imply that the Fourier transform of any such function
$V_\Gamma(\bm{p},\bm{q})$ is a configuration-space central
potential $V_\Gamma(r),$ $r\equiv|\bm{x}|.$ Splitting~off all
dependence on the angular variables then converts Salpeter's
equation (\ref{Eq:SE})~to a system of \emph{coupled\/} equations
for the radial factors of the independent components \cite{REE}.
Suppressing the index $\Gamma,$ any potential $V(r)$ enters in
these radial equations by its Fourier--Bessel~transform
$$V_L(p,q)\equiv8\pi\int\limits_0^\infty{\rm
d}r\,r^2\,j_L(p\,r)\,j_L(q\,r)\,V(r)\ ,\qquad p\equiv|\bm{p}|\
,\qquad q\equiv|\bm{q}|\ ,\qquad L=0,1,2,\dots\ ,$$given in terms
of the spherical Bessel functions of the first kind \cite{AS}
$j_n(z),$ $n=0,\pm 1,\pm2,\dots.$

Let us now zoom in onto our actual targets, i.e., the light
pseudoscalar mesons: ordinary --- in contrast to
exotic\footnote{For a fermion--antifermion bound state, its parity
$P$ and --- if this bound state is composed of a fermion and its
associated antiparticle and therefore exhibits a well-defined
behaviour under charge conjugation --- its charge-conjugation
parity $C$ are related to the relative orbital angular momentum
$\ell$ and the total spin $S$ of the bound-state constituents
according to $P=(-1)^{\ell+1}$ and $C=(-1)^{\ell+S};$ for any such
bound state with total spin $J$ conceivable quantum-number
assignments $J^{PC}$ are
$J^{PC}=0^{++},0^{-+},1^{++},1^{+-},1^{--},2^{++},\dots.$ Mesons
carrying any such assignment \emph{may\/} be quark--antiquark
bound states; their complement, those with an assignment
$J^{PC}=0^{+-},0^{--},1^{-+},2^{+-},3^{-+},4^{+-},\dots,$ must be
non-$q\overline{q}$ and belong to the \emph{exotic\/}~mesons.}
--- mesons that may be understood as bound states of a light quark
and a light antiquark. Ignoring flavour violation enforces
equality of the light-quark masses $m_1=m_2=m$ and hence of the
free energies:
$E_1(\bm{p})=E_2(\bm{p})=E(p)\equiv\sqrt{p^2+m^2},$~$p\equiv|\bm{p}|.$

In order to define unambiguously the instantaneous Bethe--Salpeter
equation we intend to invert, we need to specify the Lorentz
structure of all tensor products $\Gamma\otimes\Gamma$ of
generalized Dirac matrices $\Gamma$ in the interaction kernel
(\ref{Eq:BSK}). Following Refs.~\cite{BJK} (which form examples of
a phenomenologically acceptable analysis of the quark--antiquark
bound state spectrum),~we choose $\Gamma\otimes\Gamma$ to be the
unique (whence the right-hand side of Eq.~(\ref{Eq:BSK}) reduces
to only a single term) sophisticated linear combination of scalar,
pseudoscalar, and vector Dirac~structures\begin{equation}
\Gamma\otimes\Gamma=\frac{1}{2}\,(\gamma_\mu\otimes\gamma^\mu
+\gamma_5\otimes\gamma_5-1\otimes1)\ ,\label{Eq:LS}\end{equation}
which has the distinctive feature of Fierz symmetry, i.e.,
invariance under rearrangement of Dirac fields $\psi_k(x),$
$k=1,\dots,4$:
$\bar\psi_1(x)\,\Gamma\,\psi_2(x)\,\bar\psi_3(x)\,\Gamma\,\psi_4(x)
=\bar\psi_1(x)\,\Gamma\,\psi_4(x)\,\bar\psi_3(x)\,\Gamma\,\psi_2(x).$
With the educated guess (\ref{Eq:LS}) for the kernel, the Salpeter
equation (\ref{Eq:SE}) for spin-singlet~bound states becomes
equivalent to a set of coupled eigenvalue equations for the two
radial factors $\varphi_1(p)$ and $\varphi_2(p)$ of the
independent components of the Salpeter amplitude (\ref{Eq:PSA})
\cite[Sec.~IX]{WL07:HORSE}:
\begin{align*}&2\,E(p)\,\varphi_2(p)+2\int\limits_0^\infty\frac{{\rm
d}q\,q^2}{(2\pi)^2}\,V_0(p,q)\,\varphi_2(q)=M\,\varphi_1(p)\
,\\&2\,E(p)\,\varphi_1(p)=M\,\varphi_2(p)\ .\end{align*}The
eigenvalues of this eigenvalue problem are the possible
bound-state masses $M\equiv\sqrt{P^2}.$ The first of these coupled
relations is an integral equation encompassing all the information
about the interactions, whereas the second one is of merely
algebraic nature. For the case of interest here, i.e., for
vanishing (Goldstone-boson) mass $M=0,$ the two
relations~decouple:\begin{itemize}\item The second relation
implies that one of the Salpeter components vanishes:
$\varphi_1(p)=0.$\item The only non-zero Salpeter component,
$\varphi_2(p),$ must satisfy the bound-state equation
\begin{equation}E(p)\,\varphi_2(p)+\int\limits_0^\infty\frac{{\rm
d}q\,q^2}{(2\pi)^2}\,V_0(p,q)\,\varphi_2(q)=0\
,\label{Eq:REE}\end{equation}which, because of particularly
fortunate circumstances, is equivalent to what is called the
spinless Salpeter equation. (For reviews on the latter, consult,
e.g.,~Refs.~\cite{Lucha92,Lucha94:Como, Lucha:Oberwoelz,
Lucha:Dubrovnik,Lucha04:TWR}.)\end{itemize}

\section{Inversion Procedure}\label{Sec:IP}The actual goal of our
inversion technique \cite{WL13} is, for a given bound-state
Salpeter amplitude $\phi(\mbox{\boldmath{$p$}}),$ the extraction
of the underlying configuration-space potential $V(r)$ from the
relevant bound-state equation carved out within a sufficiently
simplified Bethe--Salpeter formalism. Preferably, this extraction
should be accomplished (as far as possible) by analytical means.

As has been shown in Ref.~\cite{WL13} by various examples, this
main goal can be easily achieved if representing, by application
of a Fourier transformation, the bound-state equation in use in
configuration space. The Fourier transformation in three
dimensions of any purely~radial function reduces to its ($L=0$)
Fourier--Bessel transformation involving the $n=0$ spherical
Bessel function of the first kind $j_0(z)=(\sin z)/z.$ Defining
the Fourier--Bessel transforms of \emph{momentum-space\/} Salpeter
component $\varphi_2(p)$ and kinetic-energy contribution
$E(p)\,\varphi_2(p)$~by\begin{align}&\varphi(r)\equiv\sqrt\frac{2}{\pi}
\int\limits_0^\infty{\rm d}p\,p^2\,j_0(p\,r)\,\varphi_2(p)\
,\qquad T(r)\equiv\sqrt\frac{2}{\pi}\int\limits_0^\infty{\rm
d}p\,p^2\,j_0(p\,r)\,E(p)\,\varphi_2(p)\ ,\label{Eq:FBT}
\end{align}Eq.~(\ref{Eq:REE}) --- the Bethe--Salpeter quintessence
relevant here --- becomes in configuration~space
$$T(r)+V(r)\,\varphi(r)=0\ .$$Thus, anticipating that $\varphi(r)$
has no zeros, for $M=0$ the central potential in question~reads
\begin{equation}V(r)=-\frac{T(r)}{\varphi(r)}\ .\label{Eq:po}
\end{equation} N.B.: From Eqs.~(\ref{Eq:FBT}) and
(\ref{Eq:po}), both $T(r)$ and $V(r)$ approach, for large $m,$
their trivial~limits
$$E(p)\xrightarrow[m\to\infty]{}m\qquad\Longrightarrow\qquad
T(r)\xrightarrow[m\to\infty]{}m\,\varphi(r)\qquad\Longrightarrow\qquad
V(r)\xrightarrow[m\to\infty]{}-m\ .$$

\section{Salpeter Amplitudes of Light Pseudoscalar Mesons}
\label{Sec:PSA}At this stage, the only ingredient to the
application of our inversion approach still lacking is the
Salpeter amplitude of the bound states in the focus of our
interest. In order to distill, by a heuristic line of argument, at
least some rough idea of the shape of the Salpeter amplitude
describing the light pseudoscalar meson, we take advantage of a
relationship,~proven within the context of Dyson--Schwinger
equations \cite{PM97a,PM97b}, between this Salpeter amplitude, on
the one hand, and the quark mass function in the dressed quark
propagator, on the~other~hand.

\subsection{Preliminary Definitions}For simplicity of notation, let
us introduce the bound-state vertex function $\Gamma(p,P),$
derived from the Bethe--Salpeter amplitude $\Phi(p,P)$ by removal
of both fermion propagators~$S_i(p_i)$:
$$\Gamma(p,P)=S_1^{-1}(p_1)\,\Phi(p,P)\,S_2^{-1}(-p_2)
\qquad\Longleftrightarrow\qquad
\Phi(p,P)=S_1(p_1)\,\Gamma(p,P)\,S_2(-p_2)\ .$$Using this quantity
where convenient, the homogeneous Bethe--Salpeter equation
(\ref{Eq:BSE}) reads$$\Gamma(p,P)=\frac{{\rm i}}{(2\pi)^4}\int{\rm
d}^4q\,K(p,q,P)\,\Phi(q,P)\ .$$

Any exact fermion propagator $S(p)$ arises as a solution of the
fermion Dyson--Schwinger (or gap) equation and, by Lorentz
covariance and parity conservation, has to be of the~form
$$S(p)=\frac{{\rm i}}{A(p^2)\!\not\!p-B(p^2)+{\rm
i}\,\varepsilon}=\frac{{\rm i}\,Z(p^2)}{\not\!p-M(p^2)+{\rm
i}\,\varepsilon}\ ,\qquad\not\!p\equiv p^\mu\,\gamma_\mu\
,\qquad\varepsilon\downarrow0\ ,$$where $A(p^2)$ and $B(p^2)$ are
two real Lorentz-scalar functions which may be reinterpreted as
the mass and wave-function renormalization functions of the
respective fermion by~defining
$$M(p^2)\equiv\frac{B(p^2)}{A(p^2)}\equiv Z(p^2)\,B(p^2)\ ,\qquad
Z(p^2)\equiv\frac{1}{A(p^2)}\ .$$

\subsection{Results from Euclidean-Space Dyson--Schwinger
Formalism}For sound reasons \cite{RLA}, the apparatus of
Dyson--Schwinger equations is typically developed in Euclidean
space, with metric $g_{\mu\nu}=\delta_{\mu\nu}.$ Hence, for
clarity, we identify in the following all Euclidean-space
variables by underlining. Ignoring overall quark-flavour factors,
the vertex function $\Gamma(\underline{k},\underline{P})$ for a
generic pseudoscalar meson P, with leptonic decay constant~$f_{\rm
P},$ reads$$\Gamma(\underline{k},\underline{P})=\gamma_5\left[{\rm
i}\,E(\underline{k},\underline{P})+\cdots\right],$$where
$E(\underline{k},\underline{P})$ is the \emph{dominant\/} Dirac
component, and the dots indicate the contributions of subleading
\cite{PM97b} Dirac components. We can achieve our goal by
combining two observations:\begin{enumerate}\item In the chiral
limit, the renormalized axial-vector Ward--Takahashi identity
relates the Dirac component $E(\underline{k},\underline{P})$ for
$\underline{P}=0$ and the quark self-energy function
$B(\underline{k}^2)$ \cite{PM97a,PM97b}:$$f_{\rm
P}\,E(\underline{k},0)=B(\underline{k}^2)\ .$$In the chiral limit,
the vertex function of a (because of its Goldstone nature)
massless ($\underline{P}^2=0$) pseudoscalar meson thus reads, in
the center-of-momentum frame ($\underline{\bm{P}}=0$),
$$\Gamma(\underline{k},0)=\gamma_5\left[{\rm
i}\,E(\underline{k},0)+\cdots\right]=\gamma_5\left[\frac{{\rm
i}}{f_{\rm P}}\,B(\underline{k}^2)+\cdots\right].$$Reinstalling
both quark propagators yields the associated Bethe--Salpeter
amplitude$$\Phi(\underline{k},0)
=S(\underline{k})\,\Gamma(\underline{k},0)\,S(\underline{k})
\propto\frac{Z(\underline{k}^2)\,M(\underline{k}^2)}
{\underline{k}^2+M^2(\underline{k}^2)}\,\gamma_5+\cdots\ .$$\item
We deduce the form of $\Phi(\underline{k},0)$ from an explicit
solution \cite{PM97b} for the quark propagator:\begin{itemize}
\item The wave-function renormalization is usually very close to
unity, i.e., $Z(\underline{k}^2)\lessapprox1.$ Letting, for
simplicity, $Z(\underline{k}^2)\approx1,$ i.e.,
$M(\underline{k}^2)\approx B(\underline{k}^2),$ yields our
starting point\begin{equation}\displaystyle\Phi(\underline{k},0)
\propto\frac{M(\underline{k}^2)}{\underline{k}^2+M^2(\underline{k}^2)}\,
\gamma_5+\cdots\ .\label{Eq:SP}\end{equation}

\newpage

\item The Dyson--Schwinger equations constitute an infinite tower of
coupled integral equations for the infinite number of $n$-point
Green functions of a given quantum field theory. This system of
equations relates every $n$-point Green function~to, at least, one
$n'$-point Green function with $n'>n.$ Accordingly, finding
solutions to such infinite hierarchy of Dyson--Schwinger equations
requires, in the first place, the formulation of a manageable
problem by truncation to a small finite number of relations for
the low-order Green functions. Each higher-order Green function
demanded as input by the truncated system of relations can only be
\emph{modelled\/}~in accordance with all its general features
expected from the quantum~field theory.

Clearly, both internal consistency and physical meaningfulness
require any such truncation to be \emph{compatible\/} with, at
least, each of the identities in the totality of Ward--Takahashi
identities encoding the symmetries of the underlying quantum field
theory that proves to be indispensable for the problem under
consideration \cite{Munczek}. A truncation scheme claimed to
preserve the axial-vector Ward--Takahashi identity is the
rainbow-ladder truncation, requiring a couple of approximations:
\begin{itemize}\item The exact (dressed) \emph{quark--gluon vertex
function\/} is replaced by its tree-level approximation; this
change is usually dubbed as ``rainbow approximation.''\item At
each occurrence, the \emph{Bethe--Salpeter kernel\/} $K$ is
reduced to (an iteration of) its lowest-order perturbative
contribution, single-gluon exchange, which results in the ``ladder
approximation'' of any Bethe--Salpeter-type equation.\item The
exact (dressed) gluon propagator is replaced by its free
approximation.\item An appropriate effective coupling replacing
the product of strong couplings in the ladder kernel thus obtained
takes care of all phenomenological issues.\end{itemize}A
``renormalization-group-improved'' variant \cite{PM97b} in the
class of rainbow--ladder truncation models is specified by two
crucial properties of the effective coupling:\begin{itemize}\item
In the infrared, $\underline{k}^2\to0,$ it shows the pronounced
enhancement suggested by solutions of the Dyson--Schwinger
equation for the exact gluon propagator.\item In the ultraviolet,
$\underline{k}^2\to\infty,$ it approaches the perturbative
behaviour of the strong fine-structure coupling and reproduces
trivially asymptotic~freedom.\end{itemize}\end{itemize}In the
chiral limit, obtaining as solution of the Dyson--Schwinger
equation for the full quark propagator nonvanishing dynamical
quark masses $M(\underline{k}^2)$ is a consequence and hence a
signal of dynamical breakdown of chiral symmetry. In that model of
Ref.~\cite{PM97b}, the one-loop behaviour of the quark mass
function $M(\underline{k}^2)$ for large Euclidean relative momenta
$\underline{k}$ may be cast into a shape that involves the
anomalous mass dimension~$\gamma_m$:
$$\lim_{\underline{k}^2\to\infty}M(\underline{k}^2)\propto
\frac{1}{\underline{k}^2\,(\log\underline{k}^2)^{1-\gamma_m}}\
,\qquad\gamma_m=\frac{12}{33-2\,N_f}\ .$$\end{enumerate}Putting
things together, the Bethe--Salpeter amplitude (\ref{Eq:SP})
behaves, in the ultraviolet,~like
$$\lim_{\underline{k}^2\to\infty}\Phi(\underline{k},0)\propto
\frac{M(\underline{k}^2)}{\underline{k}^2}\,\gamma_5\propto
\frac{1}{\underline{k}^4\,(\log\underline{k}^2)^{1-\gamma_m}}
\,\gamma_5\ .$$Ignoring the logarithmic correction, the
Bethe--Salpeter amplitude is thus characterized~by
$$\Phi(0,0)\propto\frac{1}{M(0)}\,\gamma_5\ ,\qquad
\lim_{\underline{k}^2\to\infty}\Phi(\underline{k},0)
\propto\frac{1}{\underline{k}^4}\,\gamma_5\ .$$Mimicking the
definition of the Salpeter amplitude by integration w.r.t.\
Euclidean time~$\underline{k}_4,$ we find that Salpeter amplitudes
of light pseudoscalar mesons fall off like
$|\underline{\bm{k}}|^{-3}$ for large~$|\underline{\bm{k}}|.$

\subsection{Implementation in Minkowski Space}In the course of the
above sketched expedition or raid into the jungle of Euclidean
space, we could capture, as our main loot, two hints that might be
of help in our quest for a justifiable ansatz for the potential
behaviour of the Salpeter amplitudes of light pseudoscalar
mesons.\begin{itemize}\item For large $\bm{p}$, this amplitude
$\phi(\bm{p})$ decays proportional to the inverse third
power~of~$|\bm{p}|$:$$\varphi_2(\bm{p})\propto\frac{1}{|\bm{p}|^3}
\qquad\mbox{for}\quad\bm{p}\to\infty\ .$$\item At the origin of
three-momentum space, the amplitude $\phi(\bm{p})$ can be taken to
be~finite:$$|\varphi_2(0)|<\infty\ .$$\end{itemize}

\begin{figure}[hbt]\begin{center}\begin{tabular}{cc}
\psfig{figure=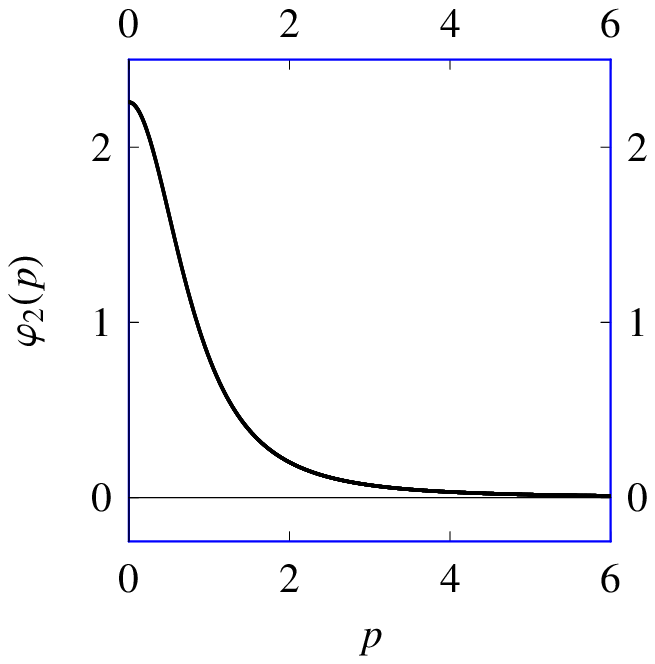,scale=1}&\psfig{figure=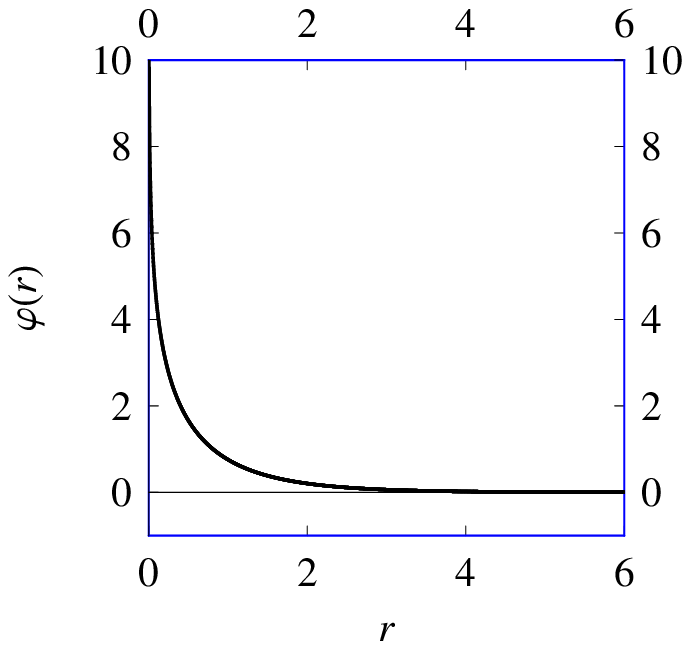,scale=1}
\\(a)&(b)\end{tabular}\caption{Salpeter amplitude (\ref{Eq:SA})
for pseudoscalar mesons: independent component relevant for, at
least, one (Fierz-invariant) Lorentz structure
$\Gamma\otimes\Gamma=\frac{1}{2}\,(\gamma_\mu\otimes\gamma^\mu
+\gamma_5\otimes\gamma_5-1\otimes1)$ of the interaction kernel in
(a) momentum space and (b) configuration space, given in units of
the parameter $\mu,$ i.e., for $\mu=1.$ In momentum-space
representation, $\varphi_2(p)=4\,(p^2+1)^{-3/2}/\sqrt{\pi}$
approaches for $p\to0$ the finite value
$\varphi_2(0)=4/\sqrt{\pi}\approx 2.256785\dots.$ In
configuration-space representation, on the other hand,
$\varphi(r)=(4\,\sqrt{2}/\pi)\,K_0(r)$ necessarily encounters for
$r\to0$ the logarithmic singularity of the modified Bessel
function $K_0(z)$: $\varphi(r)\to-(4\,\sqrt{2}/\pi)\ln(r).$}
\label{Fig:PSA-1}\end{center}\end{figure}

Any justifiable $p$ dependence of the \emph{radial\/} factor
$\varphi_2(p)$ of the Salpeter component $\varphi_2(\bm{p})$
should conform to these boundary conditions. Hence, upon
introducing a parameter $\mu$ with the dimension of mass, our hope
for analytic manageability prompts us to adopt~the~ansatz
\begin{equation}\varphi_2(p)=4\,\sqrt{\frac{\mu^3}{\pi}}\,
\frac{1}{(p^2+\mu^2)^{3/2}}\ ,\qquad\mu>0\
,\qquad\|\varphi_2\|^2\equiv\int\limits_0^\infty{\rm
d}p\,p^2\,|\varphi_2(p)|^2=1\ .\label{Eq:CMS}\end{equation}The
Fourier--Bessel transform $\varphi(r)$ of such ansatz for the
component $\varphi_2(p)$ of sole relevance for the amplitude
$\phi(\bm{p})$ is nothing but the modified Bessel function
$K_n(z)$ \cite{AS} of order~$n=0$:\begin{equation}
\varphi(r)=\frac{4\,\sqrt{2\,\mu^3}}{\pi}\,K_0(\mu\,r)\
,\qquad\mu>0\ ,\qquad\|\varphi\|^2\equiv\int\limits_0^\infty{\rm
d}r\,r^2\,|\varphi(r)|^2=1\ .\label{Eq:CSP}\end{equation}The
behaviour of this ansatz in both momentum and configuration space
is shown in~Fig.~\ref{Fig:PSA-1}. Whereas $\varphi_2(p)$ behaves
smoothly, $\varphi(r)$ diverges logarithmically at $r=0$:
$\varphi_2(0)=4/\sqrt{\pi\mu^3},$$$\varphi_2(p)
\xrightarrow[p\to\infty]{}0\ ,\qquad\varphi(r)
\xrightarrow[r\to0]{}-\frac{4\,\sqrt{2\,\mu^3}}{\pi}\ln(\mu\,r)
\xrightarrow[r\to0]{}\infty\ ,\qquad\varphi(r)
\xrightarrow[r\to\infty]{}0\ .$$

\section{Configuration-Space Radial Potential by Inversion}
\label{Sec:CSP}Our not too ambitious aim is to perform the
inversion of the Bethe--Salpeter problem~posed by the tentative
Bethe--Salpeter amplitudes (\ref{Eq:CMS}) or (\ref{Eq:CSP}) as far
as possible along an analytic path. That is to say, for as many
different values of the free quantity $m/\mu$ as manageable we
intend to derive the analytical relationship between this ansatz
for the pseudoscalar-meson solutions and the underlying
Bethe--Salpeter kernel and to extract the interaction~potential
$V(r)$ as a closed-form expression. It is straightforward to
identify (a few) choices of $m/\mu$~for which this task is easily
accomplishable. For instance, for $m=0$ (Subsec.~\ref{sec:m=0}) or
for $m=\mu$ (Subsec.~\ref{sec:m=mu}), in which case the kinetic
term $E(p)\,\varphi_2(p)$ entering $V(r)$ via $T(r)$ reduces~to
\begin{equation}E(p)\,\varphi_2(p)\propto\frac{1}{p^2+m^2}\
,\label{Eq:KT}\end{equation}since the denominator of the $p$
dependence of $\varphi_2(p)$ equals $[E(p)]^3.$ In general,
however, one has to content oneself with a numerical construction
of the potentials $V(r)$ (Subsec.~\ref{sec:marb}).\footnote{A
rather condensed preliminary account of some of the results in
this section may be found in Ref.~\cite{WLPoS}.}

\subsection{Bound-state constituents of vanishing mass: massless
quarks}\label{sec:m=0}
\begin{figure}[hbt]\begin{center}\psfig{figure=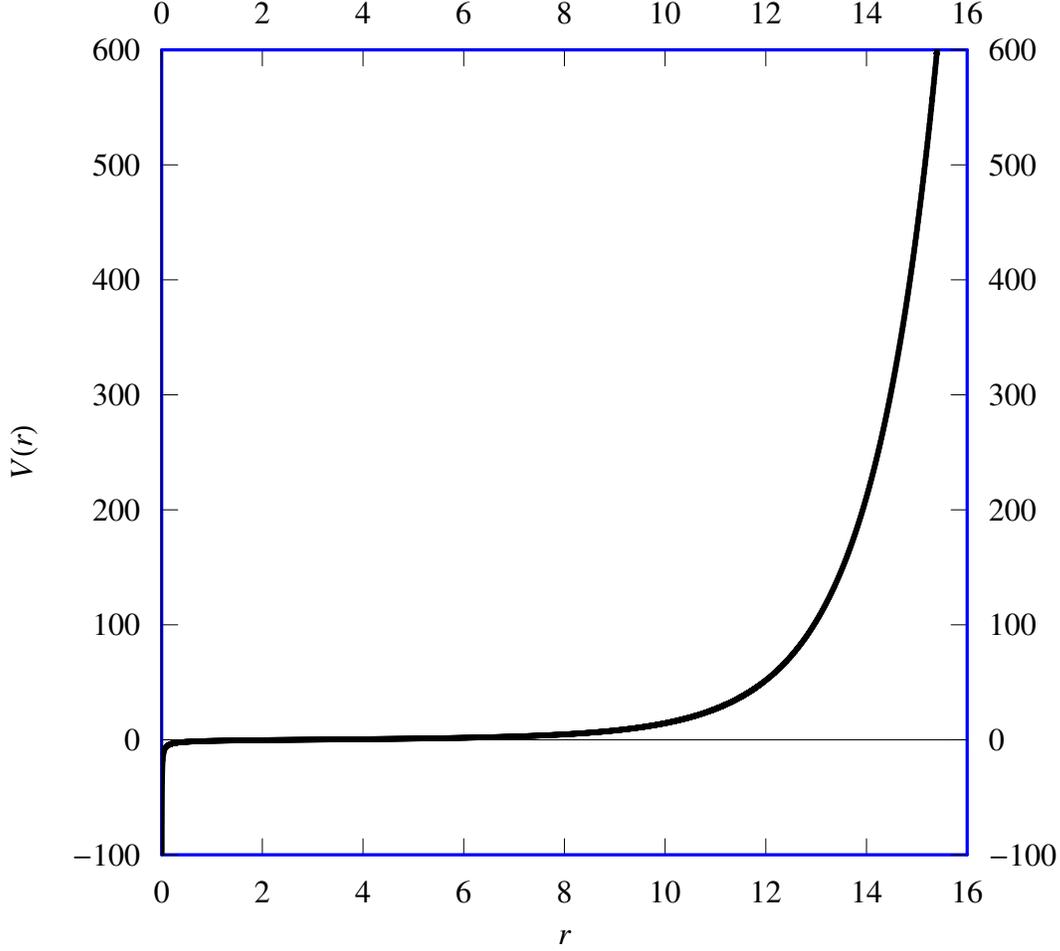,scale=1}
\caption{Configuration-space potential $V(r)$ deduced by inversion
of the Salpeter equation (\ref{Eq:SE}) with the Lorentz structure
$\Gamma\otimes\Gamma=\frac{1}{2}\,(\gamma_\mu\otimes\gamma^\mu
+\gamma_5\otimes\gamma_5-1\otimes1)$ of the interaction kernel for
the ansatz $\varphi_2(p)\propto(p^2+1)^{-3/2}$ of the relevant
component of the Salpeter amplitude $\phi(\bm{p})$ in momentum
space that is expected to describe {\em massless\/} ($M=0$)
pseudoscalar bound states of zero-mass ($m_{1,2}=0$) constituents:
$V(r)=\pi\,[r\,{\bf L}_{-1}(r)+{\bf L}_0(r)-I_0(r)-r\,I_1(r)]/
[2\,r\,K_0(r)]$ exhibits a singularity $(r\ln r)^{-1}$ at $r=0$
and a {\em confining\/} rise to infinity for large distances~$r.$}
\label{Fig:PoPSA-0}\end{center}\end{figure}

In the ultrarelativistic limit, things
become very simple. For $m=0$ and therefore $E(p)=p,$ the
Fourier--Bessel transform $T(r)$ of the kinetic term involves the
modified Bessel function $I_n(z)$ \cite{AS} (of order $n=0,1$) and
the modified Struve function ${\bf L}_n(z)$ \cite{AS} (of order
$n=-1,0$):$$T(r)=\frac{2\,\sqrt{2\,\mu^3}}{r}\left[
I_0(\mu\,r)+\mu\,r\,I_1(\mu\,r)-\mu\,r\,{\bf L}_{-1}(\mu\,r)-{\bf
L}_0(\mu\,r)\right].$$So, a potential (\ref{Eq:po}) yielding
\emph{massless\/} pseudoscalar bound states of \emph{massless\/}
constituents~is$$V(r)=\frac{\pi}{2}\,\frac{\mu\,r\,{\bf
L}_{-1}(\mu\,r)+{\bf L}_0(\mu\,r)-I_0(\mu\,r)-\mu\,r\,I_1(\mu\,r)}
{r\,K_0(\mu\,r)}\ .$$As depicted in Fig.~\ref{Fig:PoPSA-0}, this
potential is characterized by a logarithmically softened Coulomb
singularity at the origin $r=0$ and a \emph{confining\/} rise
beyond bounds for large distances
$r,$~i.e.,$$V(r)\xrightarrow[r\to0]{}\frac{\pi}{2\,r\ln(\mu\,r)}
\xrightarrow[r\to0]{}-\infty\ ,\qquad V(r)
\xrightarrow[r\to\infty]{}\sqrt{\frac{8}{\pi\,\mu^5\,r^7}}
\exp(\mu\,r)\xrightarrow[r\to\infty]{}\infty\ .$$The negative
portion of $V(r)$ for small distances $r,$ in cooperation with the
strongly peaked $r$-dependence of $\varphi(r)$ near $r=0,$ serves
to counterbalance the positive contributions to the bound-state
mass $M$ arising from the kinetic term and from the positive
portion of $V(r).$ It is therefore crucial for maintaining the
desired masslessness of the bound~state,~i.e.,~$M=0.$ In
Subsec.~\ref{sec:marb}, the above $r$ dependence will prove to
form a prototype for the behaviour of the potential $V(r)$ for any
mass $m$ of the bound-state constituents in the region $0\le
m<\mu.$

\subsection{Bound-state constituents of non-zero mass: massive
quarks}\label{sec:m>0}

\subsubsection{Case $\bm{0<m=\mu}$}\label{sec:m=mu}
\begin{figure}[hbt]\begin{center}\psfig{figure=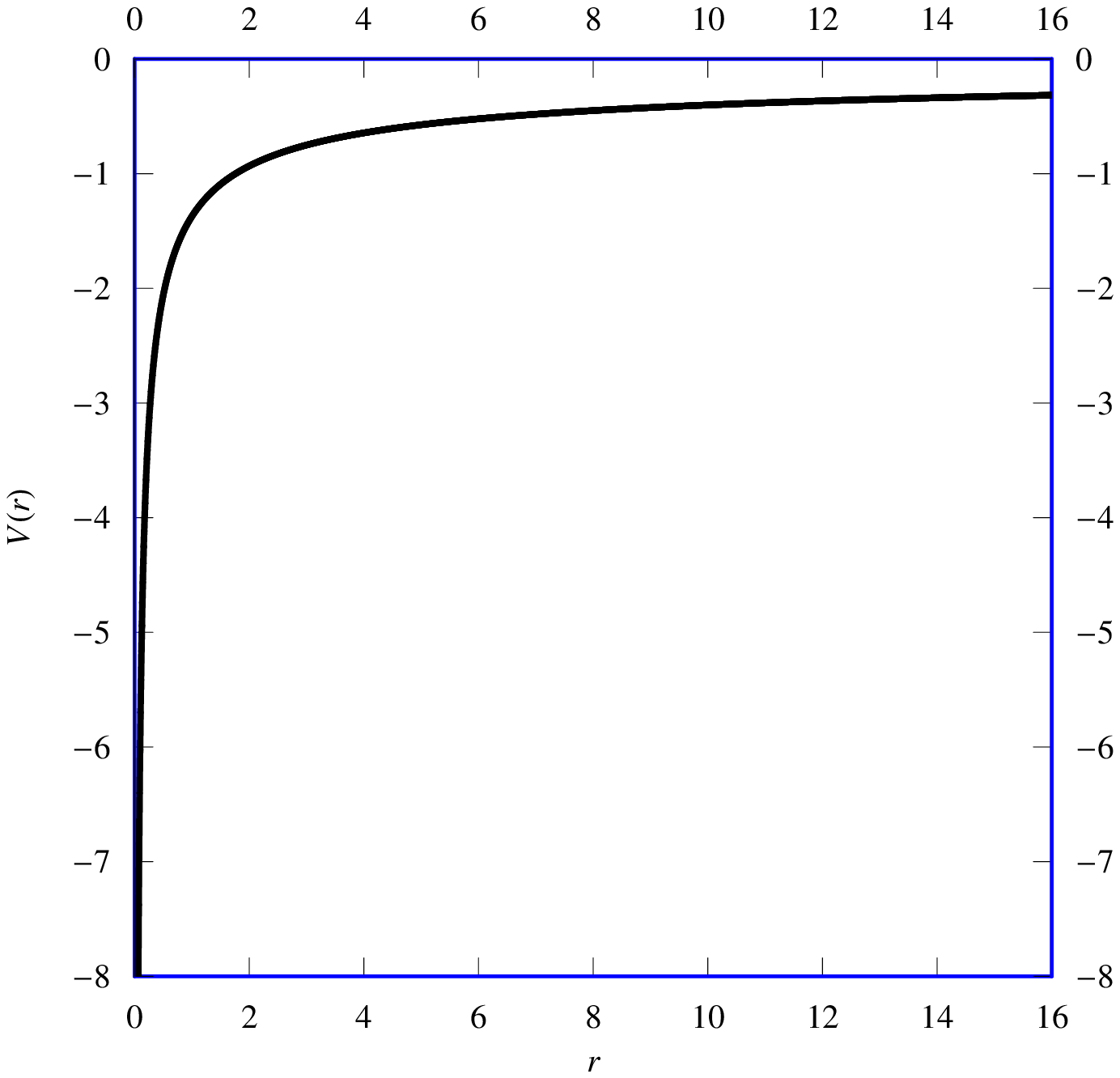,scale=1}
\caption{Configuration-space potential $V(r)$ deduced by inversion
of the Salpeter equation (\ref{Eq:SE}) with the Lorentz structure
$\Gamma\otimes\Gamma=\frac{1}{2}\,(\gamma_\mu\otimes\gamma^\mu
+\gamma_5\otimes\gamma_5-1\otimes1)$ of the interaction kernel for
the ansatz $\varphi_2(p)\propto(p^2+1)^{-3/2}$ of the relevant
component of the Salpeter amplitude $\phi(\bm{p})$ in momentum
space that is expected to describe {\em massless\/} ($M=0$)
pseudoscalar bound~states of constituents with non-vanishing
masses $m_{1,2}=1$: $V(r)=-\pi\exp(-r)/[2\,r\,K_0(r)]$ shows for
$r\to0$ the same logarithmically softened Coulomb singularity
$(r\ln r)^{-1}$ as the potential found for zero-mass constituents
(cf.\ Fig.~\ref{Fig:PoPSA-0}) but, in distinct contrast to the
behaviour of the latter, approaches for $r\to\infty$ a finite
value, $0,$ and is, accordingly, a {\em
nonconfining\/}~potential.}\label{Fig:PoPSA-mum}\end{center}
\end{figure}

If the
non-vanishing common mass $m$ of the two bound-state constituents
is precisely equal to the free ``smoothing'' parameter $\mu,$ that
is, for $0<m=\mu,$ the kinetic term $E(p)\,\varphi_2(p)$ is, from
Eq.~(\ref{Eq:KT}), proportional to the Fourier transform of the
Yukawa shape~$\exp(-z)/z;$ thus, its Fourier--Bessel transform
$T(r)$ is inevitably of Yukawa form, with $m$ as slope~parameter:
$$T(r)=\frac{2\,\sqrt{2\,m^3}}{r}\exp(-m\,r)\ .$$Accordingly,
under the circumstances discussed above, a potential that leads,
as solution of the Salpeter equation, to \emph{massless\/}
pseudoscalar bound states of \emph{massive\/} constituents~reads
$$V(r)=-\frac{\pi}{2}\,\frac{\exp(-m\,r)}{r\,K_0(m\,r)}\
.$$Similarly to the case of zero-mass constituents inspected in
Subsec.~\ref{sec:m=0}, the above potential has a logarithmically
softened Coulomb singularity at the origin $r=0,$ where the
Coulomb singularity reflects the Yukawa-type shape of $T(r),$
whereas the logarithmic softening again derives from the
(singular) behaviour of the modified Bessel function
$K_0(z\to0)\approx-\ln(z);$ in contrast, for large
inter-constituent separation $r$ the potential approaches zero
like $r^{-1/2}$:$$V(r)\xrightarrow[r\to0]{}
\frac{\pi}{2\,r\ln(m\,r)}\xrightarrow[r\to0]{}-\infty\ ,\qquad
V(r)\xrightarrow[r\to\infty]{}-\sqrt{\frac{\pi\,m}{2\,r}}
\xrightarrow[r\to\infty]{}0\ .$$Such singular $V(r)$ shape
(Fig.~\ref{Fig:PoPSA-mum}) cancels exactly the contribution of the
kinetic term $T(r);$ this kind of \emph{nonconfining\/} behaviour
will turn out to be generic for arbitrary ratio~$m/\mu\ge1.$

\subsubsection{Case $\bm{0<m}$}\label{sec:marb}
\begin{figure}[hbt]\begin{center}\psfig{figure=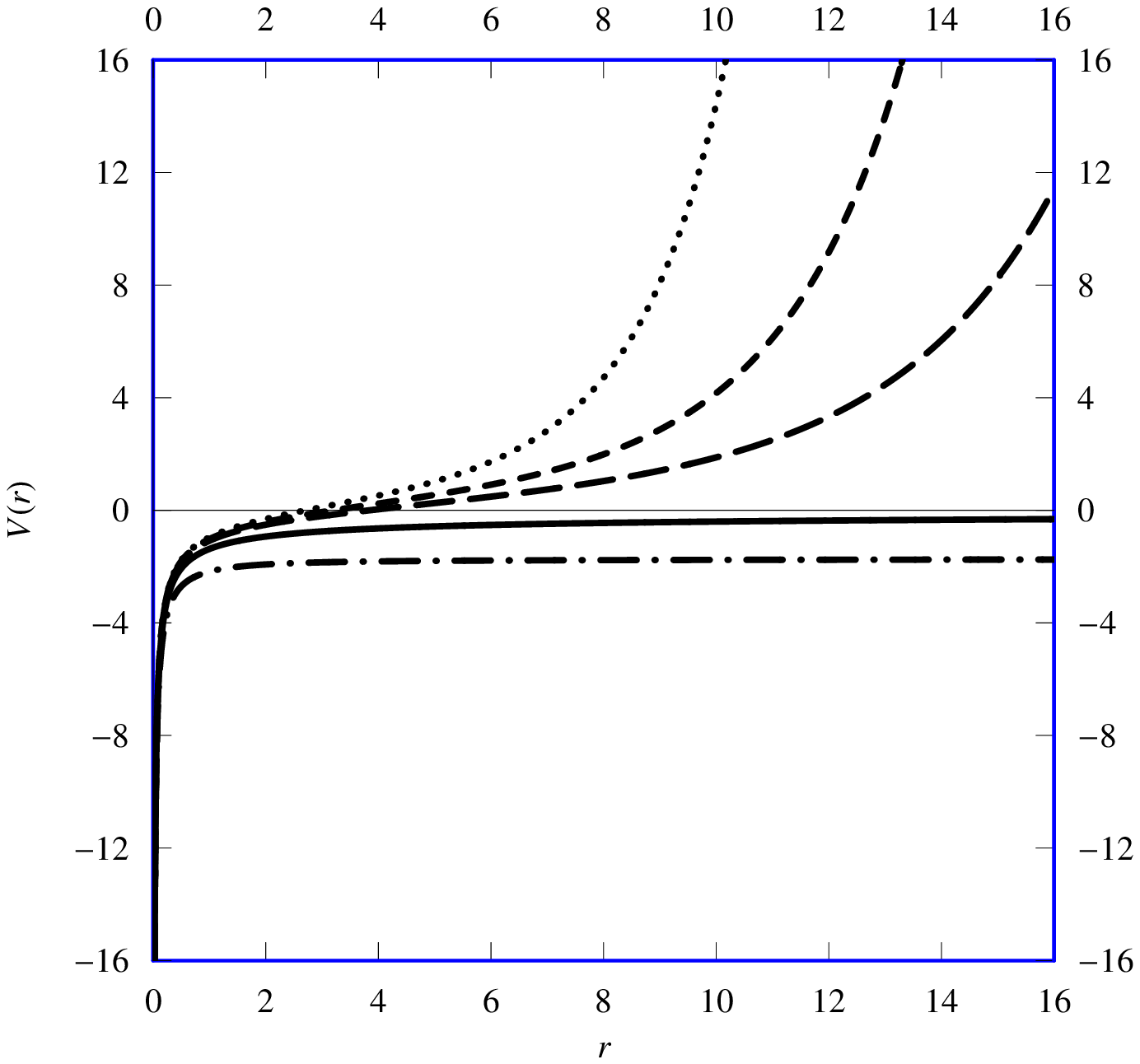,scale=1}
\caption{Configuration-space potential $V(r)$ deduced by inversion
of the Salpeter equation (\ref{Eq:SE}) with the Lorentz structure
$\Gamma\otimes\Gamma=\frac{1}{2}\,(\gamma_\mu\otimes\gamma^\mu
+\gamma_5\otimes\gamma_5-1\otimes1)$ of the interaction kernel for
the ansatz $\varphi_2(p)\propto(p^2+1)^{-3/2}$ of the relevant
component of the Salpeter amplitude $\phi(\bm{p})$ in momentum
space that is expected to describe {\em massless\/} ($M=0$)
pseudoscalar bound states of constituents with arbitrary masses
$0\le m_1=m_2\equiv m.$ In two exceptional cases, a closed
expression for $V(r)$ may be obtained: $V(r)=\pi\,[r\,{\bf
L}_{-1}(r)+{\bf L}_0(r)-I_0(r)-r\,I_1(r)]/[2\,r\,K_0(r)]$ for
$m=0$ (dotted line, see Fig.~\ref{Fig:PoPSA-0}) and
$V(r)=-\pi\exp(-r)/[2\,r\,K_0(r)]$ for $m=1$ (solid~line, see
Fig.~\ref{Fig:PoPSA-mum}); for other masses $m$ exemplifying the
general case, $m=0.35$ (short-dashed~line), $m=0.5$ (long-dashed
line), and $m=2$ (dot-dashed line), $V(r)$ has been found
numerically.}\label{Fig:PoPSA-mab}\end{center}\end{figure}

For arbitrary
non-vanishing values of the common mass $m$ of the bound-state
constituents, i.e., for $0<m\ne\mu,$ expecting the resulting
potential to be expressible in closed form~would betray an
incommensurately high degree of optimism. In general, we have to
be satisfied by extracting the potential $V(r)$ by numerical
evaluation of Eq.~(\ref{Eq:po}). In order to get~some idea about
the emerging regularities, Fig.~\ref{Fig:PoPSA-mab} depicts the
outcomes of this extraction for~a~number of selected
representative values of $m.$ Inspection of these findings leads
us to conclude that the crucial quantity determining the behaviour
of $V(r),$ at least for large separation~$r$~of the constituents,
is the relative magnitude of their mass $m$ and the wave-function
parameter~$\mu$:\begin{itemize}\item At the origin $r=0,$ the
potential $V(r)$ seemingly displays, irrespective of the value of
the constituents' mass $m,$ a logarithmically softened Coulomb
singularity of the~form$$V(r)\xrightarrow[r\to0]{}
\frac{\pi}{2\,r\ln(\mu\,r)}\xrightarrow[r\to0]{}-\infty\ .$$\item
For masses $m$ smaller than the parameter $\mu,$ $0\le m<\mu,$ the
potential $V(r)$ increases, for large $r,$ without limits, and
constitutes thus a manifest realization of confinement.\item For
masses $m$ not less than the parameter $\mu,$ $0<\mu\le m,$ the
potential $V(r)$~tends, for rising $r,$ towards a nonpositive
constant that, in the limit of large $m,$ approaches~$-m$:
$$\lim_{r\to\infty}V(r)\xrightarrow[m\to\infty]{}-m\
.$$\end{itemize}

\section{Summary of Findings}\label{Sec:SCO}Combined
Dyson--Schwinger--Bethe--Salpeter studies show that the
bound-state amplitude of a light pseudoscalar meson decreases, for
very large relative distances between the bound quark and
antiquark, approximately like a power law. In the present
analysis, we addressed the seemingly innocent question how
interaction potentials that --- when being fed into the
instantaneous Bethe--Salpeter formalism --- yield bound states
with asymptotic power-law decrease might look like. The resulting
potential exhibits presumably unexpected features. For increasing
interquark separation, the potential is monotone rising. It starts
at a sharply peaked Coulomb-like singularity at the origin but
flattens off at intermediate distances. For large distances, it
rises, for both quarks sufficiently light, rather steeply to
infinity,~whereas, for all heavier quarks, it approaches a
nonpositive finite value; the borderline for the change of
behaviour is drawn by the size of a mere model parameter which,
however, in a genuinely fundamental treatment will result from the
basic parameters of quantum~chromodynamics.

\section*{Acknowledgements}W.~L.\ is deeply indebted to Craig
D.~Roberts for several instructive and helpful~discussions.

\end{document}